\documentclass[10pt,twocolumn,superscriptaddress,english,pra,aps]{revtex4-1}
\usepackage[utf8]{inputenc}
\usepackage{amsmath}
\usepackage{amssymb}
\usepackage{graphicx}
\usepackage{xspace}
\usepackage{mathtools}
\usepackage{xcolor}
\usepackage{bbold}
\usepackage{tikz}
\usepackage{comment}
\graphicspath{{fig/}}
\makeatletter
\@ifundefined{textcolor}{}
{\definecolor{BLACK}{gray}{0}
 \definecolor{WHITE}{gray}{1}
 \definecolor{RED}{rgb}{1,0,0}
 \definecolor{GREEN}{rgb}{0,1,0}
 \definecolor{BLUE}{rgb}{0,0,1}
 \definecolor{CYAN}{cmyk}{1,0,0,0}
 \definecolor{MAGENTA}{cmyk}{0,1,0,0}
 \definecolor{YELLOW}{cmyk}{0,0,1,0}
  \definecolor{PURPLE}{cmyk}{33,67,0,40}
}

\usepackage[normalem]{ulem}

\usepackage{soul}
\usepackage{braket}
\usepackage[backref=none,
bookmarksnumbered=true,
bookmarks=true,
bookmarksopen=true,
colorlinks=true,
citecolor=blue,
linkcolor=blue,
anchorcolor=green,
urlcolor=blue,unicode=false]{hyperref}
\let\vaccent=\v % rename builtin command \v{} to \vaccent{}
\renewcommand{\v}[1]{\ensuremath{\mathbf{#1}}} % for vectors
 
\let\baraccent=\= % rename builtin command \= to \baraccent
\renewcommand{\=}[1]{\stackrel{#1}{=}} % for putting numbers above =

\makeatother

\usepackage[english]{babel}

\hypersetup{
           breaklinks=true,   % splits links across lines
           colorlinks=true,   % displays links as colored text instead of blocks
           pdfusetitle=true,  % \title and \author values into pdf metadata
        }

\begin{document}
\title{Anderson's theorem for correlated insulating states in twisted bilayer graphene}

\author{Kry\vaccent{s}tof Kol\'{a}\vaccent{r}}
\affiliation{\mbox{Dahlem Center for Complex Quantum Systems and Fachbereich Physik, Freie Universit\"at Berlin, 14195 Berlin, Germany}}

\author{Gal Shavit}
\affiliation{Department of Condensed Matter Physics,    Weizmann Institute of Science, Rehovot, Israel 7610001}

\author{Christophe Mora}
\affiliation{Universit\'e  de  Paris, Laboratoire  Mat\'eriaux  et  Ph\'enom\`enes  Quantiques, CNRS, 75013  Paris,  France}

\author{Yuval Oreg}
\affiliation{Department of Condensed Matter Physics,    Weizmann Institute of Science, Rehovot, Israel 7610001}

\author{Felix von Oppen}
\affiliation{\mbox{Dahlem Center for Complex Quantum Systems and Fachbereich Physik, Freie Universit\"at Berlin, 14195 Berlin, Germany}}

\date{\today}
\begin{abstract}
The emergence of correlated insulating phases in magic-angle twisted bilayer graphene exhibits strong sample dependence. Here, we derive an Anderson theorem governing the robustness against disorder of the Kramers intervalley coherent (K-IVC) state, a prime candidate for describing the correlated insulators at even fillings of the moir\'e flat bands. We find that the K-IVC gap is robust against local perturbations, which are odd under $\mathcal{PT}$, where $\mathcal{P}$ and $\mathcal{T}$ denote particle-hole conjugation and time reversal, respectively. In contrast, $\mathcal{PT}$-even perturbations will in general induce subgap states and reduce or even eliminate the gap. We use this result to classify the stability of the K-IVC state against various experimentally relevant perturbations. The existence of an Anderson theorem singles out the K-IVC state from other possible insulating ground states.
\end{abstract}
\pacs{%
      %73.20.-r 	%Electron states at surfaces and interfaces
			%74.25.Jb, % Superconcuctivity: Electronic structure (photoemission, etc.)
			%74.55.+v % SC: Tunneling phenomena: single particle tunneling and STM
			} % insert suggested PACS numbers in braces on next line
\maketitle 
%\maketitle must follow title, authors, abstract and \pacs

{\em Introduction.---}Twisted bilayer graphene (TBG) is recently attracting much attention as a highly tunable platform of strongly correlated electrons. Twisting two graphene sheets introduces a moir\'e lattice, which supports exceptionally flat bands for certain twist angles \cite{BistritzermacdonaldPNAS}. At these magic angles, the kinetic energy of the electrons is effectively suppressed and the system is prone to developing interaction-driven correlated phases. Corresponding experiments exhibit signatures of correlated insulators, superconductivity, nematicity, integer and fractional Chern insulators, spontaneous flavor polarization, as well as orbital ferromagnetism \cite{CaoCorrelatedInsulator, CaoUnconventionalSC, EfetovAllIntegers, YankowitzTuningMATBG, NadjPergeStrain2019,Kerelsky2019,YoungTuningSC,DiracRevivals, YazdaniRevivals, BLGscreening, hBNgoldhaberGordon, hBNyoung, EfetovBernevigPRLChernSC,Uri2020Zeldovtwistangledisorder,Xie2021,Park2021FlavorHunds}. 

A prime candidate for understanding the correlated insulating phases, which occur near even integer fillings of the moir\'e flat bands are Kramers intervalley coherent (K-IVC) states \cite{BultnickKhalaf2020,Ochoa2020,hofmann2021fermionic,ShavitMaATBGprl,bultinckKwanKekulSpiralOrder2021,bernevigCalugaruSpectroscopyTwistedBilayer2021,zaletelHongDetectingSymmetryBreaking2021}. These states exhibit a pattern of magnetization currents, which triple the graphene unit cell, thereby breaking the lattice translation symmetry as well as time reversal.
The associated spontaneous coherence between the two valleys of the TBG band structure gaps out the moir\'e Dirac points and induces insulating behavior. A recent work reports evidence for the K-IVC state by measuring the magnetic-field dependence
of the thermodynamic gap \cite{feldmanYuSpinSkyrmionGaps2022}. In general, the appearance and strength of insulating states tend to be device dependent
\cite{YoungTuningSC,EfetovTuningSC}. A possible explanation for this sample-specific behavior lies in residual disorder associated with random strain or impurity potentials. Thus, it is important to study and understand their effects.  

Previous works \cite{pixleyWilsonDisorderTwistedBilayer2020,AliceaDisorder,bultinckKwanKekulSpiralOrder2021,EntropyDomains,hBNdisorderMacdonald}
considered smooth disorder and twist-angle variations, for which the associated long-range domain patterns can be directly mapped in experiment \cite{NadjPergeStrain2019,Kerelsky2019,DiracRevivals,YazdaniRevivals,Uri2020Zeldovtwistangledisorder,BocarslyZeldovChernMosaic}. Here, we focus on local impurities. We show that for the K-IVC state, one can systematically classify impurities according to their ability to induce subgap excitations, which diminish or even eliminate the insulating gap. Our discussion is strongly informed by a far-reaching analogy with the familiar problem of classifying impurities in $s$-wave superconductors \cite{Anderson1959}. 

Physically, these analogies can be understood by noting that K-IVC states can be thought of as binding holes in one valley to electrons in the other, akin to excitonic insulators. Evidently, this is similar to binding time-reversed electrons into Cooper pairs. According to Anderson's theorem \cite{Anderson1959}, the ability of impurities to induce subgap excitations in $s$-wave superconductors is controlled by whether or not they respect time-reversal symmetry. We find that particle-hole symmetry plays a similar role for the K-IVC state as time-reversal symmetry does for superconductors. This is consistent with the picture of bound electron-hole pairs.

In contrast to the K-IVC state, we find that there are no corresponding Anderson theorems for other possible insulating ground states such as the valley polarized and valley Hall states.

{\em Anderson's theorem for $s$-wave superconductors.---} To clearly bring out the analogies, as well as differences, between TBG and superconductors, we begin our discussion with a review of Anderson's theorem for $s$-wave superconductors
\cite{Anderson1959,maki1969gapless,Timmons2020,andoAndersenGeneralizedAndersonTheorem2020}, using a formulation which turns out to be adaptable to TBG. Starting from the second-quantized BCS mean-field Hamiltonian $\mathcal{H} = \frac{1}{2}\int d\mathbf{r}\Psi^\dagger(\mathbf{r}) H  \Psi(\mathbf{r})$, we write the Bogoliubov-de Gennes (BdG) Hamiltonian
\begin{equation}
   H = \left( \begin{array}{cc}  H_e & \Delta \\ \Delta & H_h 
   \end{array}\right),
   \label{Hsup}
\end{equation}
in a four-component Nambu formalism, using the basis $\Psi = [\psi^{\phantom{\dagger}}_\uparrow,\psi^{\phantom{\dagger}}_\downarrow,\psi^\dagger_\downarrow,-\psi^\dagger_\uparrow]$. The normal-state Hamiltonians $H_{e/h}$ for electrons ($e$) and holes ($h$) as well as the pairing $\Delta$ are matrices in spin space. For $s$-wave pairing, $\Delta$ is proportional to the unit matrix. In the following, we assume that $\Delta$ is spatially uniform and chosen to be real. 

The BdG Hamiltonian is constrained by antisymmetry under particle-hole conjugation, $\mathcal{P}H\mathcal{P}^{-1} = -H$. As a consequence, the normal-state Hamiltonians of electrons and holes in Eq.\ (\ref{Hsup}) are related by time reversal $\mathcal{T}$,
\begin{equation}
   H_h= - \mathcal{T} H_e \mathcal{T}^{-1}.
   \label{normalHT}
\end{equation}
This can be deduced by defining Pauli matrices $\tau_\alpha$ and $s_\alpha$ in particle-hole and spin space, respectively. Then, particle-hole conjugation is implemented by 
$\mathcal{P}=-i\tau_y \mathcal{T}=\tau_y s_y \mathcal{K}$ and squares to unity, $\mathcal{P}^2=1$, while time reversal takes the form $\mathcal{T}=i s_y \mathcal{K}$ with $\mathcal{T}^2=-1$ ($\mathcal{K}$ implements complex conjugation).

We separate the normal-state Hamiltonian $H_e=H_0+U$ into a spatially homogeneous part $H_0$ and a (local) perturbation $U$. While we assume $H_0=\mathcal{T}H_0\mathcal{T}^{-1}$ to be time-reversal symmetric, a general  perturbation $U = U_+ + U_-$ can have components $U_\pm = \pm \mathcal{T}U_\pm\mathcal{T}^{-1}$, which are even ($+$) or odd ($-$) under time reversal. Combining these symmetry properties under time reversal with Eq.\ (\ref{normalHT}), the BdG Hamiltonian in Eq.\ (\ref{Hsup}) can be written compactly as 
\begin{equation}
   H = H_0\tau_z + \Delta\tau_x  + U_+ \tau_z + U_- \tau_0. 
   \label{eq:compactH}
\end{equation}
Importantly, one observes that time-reversal-symmetric perturbations anticommute with the order-parameter term, $\{\Delta\tau_x,U_+ \tau_z\}=0$, while the time-reversal breaking term, $U_-$, commutes.
 
It can now be seen quite generally that time-reversal-even perturbations do not reduce the BdG gap (Anderson's theorem). Given that antisymmetry under particle-hole conjugation $\mathcal{P}$ enforces the eigenenergies to be symmetric about zero energy, the spectrum can be deduced from the square of $H$,
\begin{equation}
\label{eq:Hsquared}
H^2 = (H_0 + U_+)^2 + \Delta^2,
\end{equation}
implying that the magnitude of the eigenvalues of $H$ is bounded from below by $\Delta$. This argument uses the assumption that the gap remains uniform in the presence of the perturbation, but holds regardless of the particular spatial structure of the impurity potential.

Conversely, perturbations, which are odd under time reversal generally reduce the gap. A uniform Zeeman field described by $U_-=\mathbf{B}\cdot\mathbf{s}$ reduces the gap to $\Delta-|\mathbf{B}|$, provided the normal-state Hamiltonian is spin-rotation invariant. Local magnetic impurities with $U_-=J\mathbf{S}\cdot \mathbf{s}\delta(\mathbf{r})$ are well-known to induce Yu-Shiba-Rusinov states at subgap energies \cite{Yu1965,Shiba1968,Rusinov1969,Balatsky2006}.

{\em Twisted bilayer graphene.---}We begin our discussion of TBG by introducing its band Hamiltonian $h(\mathbf{k})$, after projection to the eight flat bands. It is conveniently written in the Chern basis spanned by the spin, valley (Pauli matrices $\tau_\alpha$), and band (Pauli matrices $\sigma_\alpha$) degrees of freedom  \cite{BultnickKhalaf2020},
\begin{equation}
\label{eq:hkinchernbasisfirst}
h(\mathbf{k}) =  h_0(\mathbf{k}) \tau_z + h_x(\mathbf{k}) \sigma_x + h_y(\mathbf{k}) \sigma_y \tau_z.
\end{equation}
Time-reversal symmetry enforces $h_0(\mathbf{k}) = -h_0(-\mathbf{k})$ and $h_{x,y}(\mathbf{k}) = h_{x,y}(-\mathbf{k})$.
Due to the negligible spin-orbit coupling of graphene, the Hamiltonian is a unit matrix in spin space. The label $\sigma_z$ is associated with the Chern number $C=\sigma_z\tau_z$ and (partial) sublattice polarization. While the Pauli matrices $\tau_\alpha$ refer to different degrees of freedom in our descriptions of superconductors and TBG, we shall see that they actually play rather analogous roles. 

In addition to spatial and spin rotation symmetries as well as charge conservation, the Hamiltonian $h(\mathbf{k})$ conserves valley charge, obeys spinless time-reversal symmetry, and has an (approximate) particle-hole antisymmetry \cite{Kang2018,Song2019,BultnickKhalaf2020,lianBernevigTBGIIIInteracting2021}. The latter three symmetries are central to our discussion. The conservation of valley charge is associated with invariance under $U(1)$ valley rotations $U_V = e^{i \theta \tau_z }$, spinless time reversal is implemented by $\mathcal{T} = \tau_x \mathcal{K}$ with $\mathcal{T}^2 = 1$, and particle-hole conjugation takes the form $\mathcal{P} = i \sigma_y \tau_z \mathcal K$ with $\mathcal{P}^2=-1$. The particle-hole antisymmetry of TBG emerges when neglecting the small relative twist of the Dirac Hamiltonians of the two layers in the Bistrizer-MacDonald model \cite{BistritzermacdonaldPNAS}.

Within the mean-field approximation for the K-IVC state \cite{BultnickKhalaf2020}, the band Hamiltonian $h(\mathbf{k})$ is complemented by the order parameter $h_{\text{IVC}} =\Delta \sigma_y (\tau_x \cos \theta + \tau_y \sin \theta )$, where $\theta$ denotes an arbitrary phase. In view of the associated magnetization currents, the K-IVC state breaks time reversal spontaneously, $\mathcal{T}h_{\text{IVC}}\mathcal{T}^{-1} = -h_{\text{IVC}}$. However, it preserves a modified time-reversal symmetry \cite{BultnickKhalaf2020}
\begin{equation}
\mathcal{T'} = i \tau_y \mathcal{K},
\end{equation}
which concatenates $\mathcal T$ with a valley rotation, $\mathcal{T'} = \tau_z \mathcal{T}$. Both $\mathcal T$ and valley rotations are symmetries of the single-particle Hamiltonian, so that the mean-field Hamiltonian $H(\mathbf{k})=h(\mathbf{k}) + h_{\text{IVC}}$ conserves the Kramers time reversal $\mathcal T'$ (with $\mathcal{T}'^2=-1$) as a whole. 

The Hamiltonian $H(\mathbf{k})$ can be thought of as the analog of the BdG Hamiltonian for the K-IVC state. We will now make the analogies yet more explicit by a change of basis $H\to \mathcal{U} H \mathcal{U}^\dagger$ with
\begin{equation}
  \mathcal{U} = \begin{pmatrix} 1 & 0\\ 0 &  i\sigma_y \end{pmatrix}.
  \label{eq:unitrans}
\end{equation}
In the new basis, which we refer to as the particle-hole basis,
the Chern number becomes $C=\sigma_z$. Transforming the Hamiltonian in this manner, we find 
\begin{equation}
   H(\mathbf{k}) = H_0(\mathbf{k})\tau_z + 
       \Delta(\tau_x \cos\tilde\theta + \tau_y \sin\tilde\theta) 
      \label{eq:HKIVC}
\end{equation}
($\tilde\theta = \theta + \frac{\pi}{2}$). Here, we make the dependence on the valley Pauli matrices $\tau_\alpha$ explicit, while $H_0$ and $\Delta$ are still matrices in sublattice space. We find $H_0(\mathbf{k})=h_0(\mathbf{k})  + h_x(\mathbf{k}) \sigma_x + h_y(\mathbf{k}) \sigma_y$ for the single-particle Hamiltonian of the $K$ valley, while $\Delta$ is simply proportional to the unit matrix. The transformation (\ref{eq:unitrans}) to the particle-hole basis also changes the explicit forms of the time-reversal and charge-conjugation operations, $\mathcal{T}\to \mathcal{U} \mathcal{T} \mathcal{U}^\dagger$ and $\mathcal{P}\to \mathcal{U} \mathcal{P} \mathcal{U}^\dagger$, which yields $\mathcal{P}= i \sigma_y \mathcal{K}$ and $\mathcal{T}'= - \tau_x \mathcal{P}$. 

Equation (\ref{eq:HKIVC}) is closely analogous to the BdG Hamiltonian of $s$-wave superconductors, with particle-hole space replaced by the valley degree of freedom and spin space replaced by sublattice space. In Eq.\ (\ref{eq:HKIVC}), the band Hamiltonian $H_0(\mathbf{k})$ is analogous to the normal-state Hamiltonian. It multiplies $\tau_z$ as a consequence of the chiral antisymmetry $\mathcal P \mathcal T = i \tau_y$ of the TBG Hamiltonian. Moreover, the term describing K-IVC order is analogous to the pairing term in the BdG Hamiltonian, being offdiagonal in valley and proportional to the unit matrix in sublattice space. 

Beyond the structural similarities of the Hamiltonians, there is also a correspondence of symmetries. Interestingly, the roles of time reversal and particle-hole conjugation are essentially reversed. For superconductors, time reversal acts diagonally in particle-hole space, while particle-hole conjugation is offdiagonal. In contrast, in TBG it is time reversal which maps between the two valleys, while particle-hole conjugation acts separately within each valley. 

We also note that gauge transformations for superconductors are structurally analogous to U(1) valley rotations in TBG. For superconductors, the BdG Hamiltonian becomes time-reversal symmetric by choosing a gauge, in which $\Delta$ is real. In TBG, we can similarly exploit the valley rotation symmetry to choose $\tilde\theta=\frac{\pi}{2}$, so that $H(\mathbf{k})=H_0(\mathbf{k})\tau_z +\Delta\tau_y$. 
With this choice, the mean-field K-IVC order is also odd under particle-hole conjugation, so that $\mathcal{P}H(\mathbf{k})\mathcal{P}^{-1}=-H(-\mathbf{k})$. In the following, we make this choice for definiteness. However, just as  Anderson's theorem for $s$-wave superconductors is not specific to a particular gauge, Anderson's theorem for TBG is not limited to this choice. 

{\em Impurities and K-IVC states.---}Armed with this far-reaching correspondence between the BdG Hamiltonian of $s$-wave superconductors and the K-IVC state of TBG, we now turn to discussing the effects of impurities on the K-IVC state. We consider impurity potentials which are sufficiently smooth on the scale of the atomic lattice, so that they preserve the $U(1)$ valley symmetry. Intervalley scattering can then be neglected and the impurity potential is diagonal in valley space. With this assumption, the low-energy Hamiltonian in the presence of an impurity potential becomes
\begin{equation}
\label{eq:compactH_TBG}
H = H_0 \tau_z + \Delta \tau_y + U_-\tau_z + U_+ \tau_0.
\end{equation}
Just as for superconductors, the impurity potentials $U_\pm$ (which are matrices in sublattice and spin space) are distinguished by their symmetry properties. For superconductors, antisymmetry under particle-hole conjugation is built into the BdG formalism. For this reason, it was sufficient to classify perturbations according to their behavior under time reversal. In contrast, for TBG, 
both Kramers time reversal and particle-hole conjugation are physical symmetries of the  Hamiltonian. Consequently, we now classify perturbations according to their transformation properties under the combined chiral symmetry operation $\mathcal{PT}=i\tau_y$, namely $(\mathcal{PT})U_+\tau_0 (\mathcal{PT})^{-1} = U_+ \tau_0$ and $(\mathcal{PT})U_-\tau_z (\mathcal{PT})^{-1} = - U_-\tau_z$. (Notice that due to valley rotation symmetry, the impurity terms transform in the same way under $\mathcal{PT}$ and $\mathcal{PT'}$.) By comparing with the discussion of Eq.\ (\ref{eq:compactH}), we can now formulate an Anderson's theorem for TBG, our central result: The gap of K-IVC states is robust against valley-preserving perturbations, which are odd under the combined chiral symmetry operation $\mathcal{PT}$. In fact, perturbations which are odd under $\mathcal{PT}$ anticommute with the K-IVC order $\Delta \tau_y$ and cannot reduce the gap. In contrast, similar to time-reversal-breaking impurities in superconductors, perturbations which are even under $\mathcal{PT}$ can induce subgap states in TBG. For perturbations that are local on the moir\'e scale, this follows as for time-reversal-breaking impurities in $s$-wave superconductors. A finite density of $\mathcal{PT}$-even impurities can thus suppress or even destroy the K-IVC gap.

So far, our discussion relied on a close structural analogy between the BdG Hamiltonian of $s$-wave superconductors and the low-energy Hamiltonian of TBG with K-IVC order. More fundamentally, the appearance of an Anderson's theorem in both theories is rooted in the fact that up to $U(1)$ rotations which leave the normal-state Hamiltonian invariant, their order parameters are proportional to a natural antisymmetry of the model, namely $\mathcal{PT}$. For both superconductors and the K-IVC state of TBG, we have $\Delta\tau_y = -i\Delta\mathcal{PT} $. Up to a gauge transformation (superconductors) or a $U(1)$ valley rotation (TBG),  
this is  equivalent to the general order-parameter term $\Delta (\tau_x \cos\theta + \tau_y \sin\theta)$. This form of the order parameter has two important consequences. First, the order-parameter and single-particle terms in the Hamiltonian anticommute, so that the single-particle energies and the order parameter add in quadrature in the mean-field excitation spectrum.
Second, this property persists in the presence of disorder, as long as the latter is odd under $\mathcal{PT}$, which is Anderson's theorem. We emphasize that Anderson's theorem does not rely on a specific symmetry of the order parameter, but only on the fact that the order parameter is proportional to a natural unitary antisymmetry of the system in the absence of the spontaneous symmetry breaking. 

One should remember that the derivation of Anderson's theorem relies on several assumptions. In particular, one assumes that the order parameter remains spatially uniform and is momentum independent. Similar to anisotropic superconductors, the order parameter of TBG exhibits some momentum dependence  \cite{BultnickKhalaf2020}. In the presence of momentum dependence, there will be no systematic anticommutation behavior between the order-parameter term and the impurity potential. Then,  Anderson's theorem no longer applies in the strict sense, and implies only enhanced, but not full protection of the gap. 
\begin{table}[t]
\centering
\begin{tabular}{ ||c|c|c|| } 
 \hline
 perturbation & matrix structure & Anderson \\ 
 \hline \hline
 layer-even pot. & $\mathbb{1}$ & $\times$ \\ 
 layer-odd pot. & $\mu_z$ & \checkmark \\ 
 layer-even sublattice pot. & $\sigma_z$ & \checkmark \\ 
 layer-odd sublattice pot. & $\sigma_z \mu_z$ & $\times$ \\
layer-even vector pot. & $\sigma_x, \sigma_y \tau_z$ & $\times$ \\ 
 layer-odd vector pot. & $\sigma_x \mu_z, \sigma_y \tau_z\mu_z$ & \checkmark \\ 
 tunneling disorder & see text & \checkmark \\ 
  \hline
\end{tabular}
\caption{Table of time-reversal symmetric perturbations (left column) and their effect on the K-IVC gap. The central column gives the matrix structure in the microscopic graphene basis of TBG, where $\mu_\alpha$, $\sigma_\alpha$ and $\tau_\alpha$ are Pauli matrices in layer, sublattice and valley space, respectively. The right column indicates the validity of Anderson's theorem. The K-IVC gap is protected against perturbations, for which Anderson's theorem is valid. Notice that in this table, we only consider strain-induced vector potentials.}
\label{tab:Anderson}
\end{table}

{\em Classifying physical perturbations.---}So far, we have phrased our discussion in rather general terms, largely relying on symmetry properties of the TBG flat bands. We now classify perturbations according to their symmetry and tabulate the presence or absence of Anderson's theorem in Table \ref{tab:Anderson}. For a given behavior under time reversal, it is sufficient to consider their transformation properties under $\mathcal{P}$, which acts separately within each valley. 

Usually, we do not know the form of perturbations in the flat-band (i.e., Chern or particle-hole) bases, but
rather in the microscopic graphene basis. Within the Bistrizer-MacDonald model \cite{BistritzermacdonaldPNAS}, the Hamiltonian $H_0$ is valley diagonal and takes the form
\begin{equation}
   H=\left(\begin{array}{cc} v_D \boldsymbol{\sigma}\cdot (\frac{1}{i} \boldsymbol{\nabla} + \mathbf{A}_t) + \phi_t& T(\mathbf{r}) \\ T^\dagger (\mathbf{r}) & v_D \boldsymbol{\sigma}\cdot (\frac{1}{i} \boldsymbol{\nabla} + \mathbf{A}_b) + \phi_b \end{array}\right) 
   \label{eq:BisMD}
\end{equation}
for the K-valley. Here, the diagonal and off-diagonal blocks are intra- and inter-layer terms, respectively, and the $\sigma_\alpha$ refer explicitly to the graphene sublattice. Potential disorder introduces layer- and sublattice-dependent potentials $\phi_{t/b}(\mathbf{r})$. Modulations in the interlayer distance cause variations of the interlayer tunneling terms $T(\mathbf{r})$. Strain introduces vector potentials $\mathbf{A}_{t/b}(\mathbf{r})$ and modifies $T(\mathbf{r})$ \cite{StrainDesignFu,parker2020straininducedZalatel,balentsBalentsGeneralContinuumModel2019}. In terms of the strain-induced displacements $\mathbf{u}_l(\mathbf{r})$ relative to the uniformly twisted bilayer, 
the components of the vector potential take the form \cite{parker2020straininducedZalatel,balentsBalentsGeneralContinuumModel2019} 
$ (A_l)_\mu =  \mathbf{K} \cdot \partial_\mu \mathbf{u}_l +   \frac{\beta\sqrt{3}}{2a} \left(u_{l,xx}-u_{l,yy},
-2 u_{l,xy}\right)$. Here, $\beta$
characterizes the sensitivity of the hopping amplitude to strain-induced displacements and $u_{l,ij}$ is the strain tensor of layer $l$. Time-reversal symmetry implies that strain-induced vector potentials are odd in valley space, while the scalar potentials are even.

In the microscopic graphene basis of the Bistritzer-MacDonald Hamiltonian in Eq.\ (\ref{eq:BisMD}), particle-hole conjugation takes the form \cite{BultnickKhalaf2020} 
\begin{equation} \label{eq:phmic}
  \mathcal{P} =i\sigma_x \mu_y \mathcal{K},
\end{equation}
where the $\mu_\alpha$ are Pauli matrices in layer space ($\mathcal{PT} =i\tau_x \sigma_x \mu_y$). The validity of Anderson's theorem for various perturbations is now readily established and tabulated in Table \ref{tab:Anderson} for time-reversal symmetric perturbations. A sublattice-symmetric potential will commute with $\mathcal{P}$, if it is layer symmetric, and anticommute with $\mathcal{P}$, if it is odd under layer exchange. According to our considerations, we find that layer-symmetric potentials induce subgap states within the K-IVC gap, but layer-odd potentials leave the K-IVC gap intact. These conclusions are reversed for sublattice-odd potentials. Tunneling disorder corresponds to a local variation in the strength of the interlayer tunneling amplitudes and thus in the parameters entering $T(\mathbf{r})$. Consequently, tunneling disorder inherits the $\mathcal P$ transformation properties of $H_0$ and Anderson's theorem applies. Finally, a layer-even vector potential (homostrain) is even under $\mathcal{P}$, while a layer-odd vector potential (heterostrain) is odd. We therefore find that Anderson's theorem applies to (local) heterostrain only.

{\em Other insulating ground states.---}While it has been argued that the K-IVC is the most favorable ground state at even integer fillings of the flat bands of TBG \cite{BultnickKhalaf2020,hofmann2021fermionic,ShavitMaATBGprl,OxfordMATBG}, other competing symmetry-broken states can also be considered \cite{macdonaldXieNatureCorrelatedInsulator2020,vishwanathLiuNematicTopologicalSemimetal2021, TBG4Exactinsulatorgroundphasediagram,SenthilAshvin2018PRX}.
We find that the K-IVC is distinct from other possible ground states due to the existence of Anderson's theorem for $\mathcal P \mathcal T$-antisymmetric disorder. It is thus conceivable that such kinds of disorder stabilize the K-IVC state relative to  competing states. 

First, consider an alternative, time-reversal-preserving intervalley coherent state, termed T-IVC. In the particle-hole basis, this state has the order parameter  
$\Delta\sigma_z(\tau_x\cos\theta + \tau_y \sin\theta)$. The T-IVC gap anticommutes with only one of the three terms of the flat-band Hamiltonian, precluding the derivation of an Anderson's theorem. 

The valley-polarized state with order parameter $\Delta\tau_z$ leads to the mean-field Hamiltonian
\begin{equation}
    H_{\rm vp} = H_0\tau_z +\Delta \tau_z
    + U_-\tau_z + U_+ \tau_0 
\end{equation}
in the particle-hole basis. The order-parameter term commutes with the band Hamiltonian, so that a gap emerges only when $\Delta$ shifts the flat bands of the two valleys sufficiently far apart in energy. The impurity problem can be considered separately for the two bands and regardless of impurity type, there is no robustness due to an Anderson's theorem. 

Finally, we consider the valley Hall state with mean field Hamiltonian
\begin{equation}
    H_{\rm vh} = H_0\tau_z +\Delta \sigma_z\tau_z  + U_-\tau_z + U_+ \tau_0
\end{equation}
in the particle-hole basis. The order-parameter term anticommutes with the flat-band Hamiltonian $H_0\tau_z$ [see Eq.\ (\ref{eq:HKIVC})] only in the chiral limit, where $h_0(\mathbf{k})=0$ \cite{Tarnopolsky2019}. In this idealized (but experimentally remote) limit, the gap is robust against perturbations, which are purely offdiagonal in sublattice space, e.g., strain disorder. 

{\em Conclusions.---}We have shown that an Anderson-type theorem protects the gap of the K-IVC state, a prime candidate for the observed robust correlated insulators of TBG, from certain types of disorder. Similar to $s$-wave superconductivity, which is robust against time-reversal preserving disorder, we find that the K-IVC gap is robust against disorder, which is odd under $\mathcal{PT}$. The robustness against some types of disorder distinguishes the K-IVC state from other candidate ground states for the correlated insulators and arises due to the special nature of the order parameter, which has the same matrix structure, up to a $U(1)$ valley rotation, as the $\mathcal P \mathcal T$ chiral antisymmetry.

This special structure of the order-parameter term also underlies the close analogy of KIVC states with  $s$-wave superconductors. In both cases, this structure guarantees that the band Hamiltonian as well as $\mathcal{PT}$-odd disorder potentials anticommute with a uniform and momentum-independent order-parameter term. For superconductors, $\mathcal{P}$ is an inherent antisymmetry, so that Anderson's theorem applies to $\mathcal{T}$-even perturbations. For TBG, both $\mathcal{P}$ and $\mathcal{T}$ are physical symmetries, so that Anderson's theorem applies to all perturbations which are odd under $\mathcal{PT}$, regardless of their behavior under $\mathcal{P}$ and $\mathcal{T}$ individually.

Our theoretical considerations can be tested by introducing impurity potentials with different behavior under $\mathcal{PT}$. Local $\mathcal{PT}$-even perturbations will in general induce subgap states, which can be probed directly using scanning tunneling microscopy. One expects that a finite density of impurities can reduce or even eliminate the K-IVC gap. We leave a detailed study of this last situation to a separate publication \cite{Shavit2022}. 

\begin{acknowledgments}
We gratefully acknowledge funding by Deutsche Forschungsgemeinschaft through CRC 183 (project C02; YO and FvO), a joint ANR-DFG project  (TWISTGRAPH; CM and FvO), 
the European Union’s Horizon 2020 research and innovation program (Grant Agreement LEGOTOP No. 788715; YO), the ISF Quantum Science and Technology program (2074/19; YO), and a BSF and NSF grant (2018643; YO).
\end{acknowledgments}

%\bibliographystyle{apsrev4-2}
%\bibliography{library}


\end{document}